\documentclass[pra,showpacs,nofootinbib]{revtex4}
\usepackage{hyperref}
\usepackage{amsmath}
\usepackage{amssymb}
\usepackage{amsthm}
\usepackage{graphics}
\usepackage{graphicx}
\usepackage{color}
\usepackage[usenames,dvipsnames,svgnames,table]{xcolor}

\long\def\ca#1\cb{}

\def\dyad#1#2{|#1\rangle\langle#2|}

\def\ket#1{|#1\rangle }

\def\AC{{\cal A}}
\def\BC{{\cal B}}
\def\CC{{\cal C}}
\def\DC{{\cal D}}

\def\HC{{\cal H}}

\def\KC{{\cal K}}

\def\SC{{\cal S}}

\def\endproof{{\hspace{\stretch{1}}$\blacksquare$}}

\newtheorem{thm1}{Theorem}
\newtheorem{cor1}[thm1]{Corollary}

\newtheorem{lem1}[thm1]{Lemma}
\newtheorem{lem2}[thm1]{Lemma}
\newtheorem{lem3}[thm1]{Lemma}
\newtheorem{lem4}[thm1]{Lemma}
\newtheorem{thm2}[thm1]{Theorem}

\begin{document}
\title{Local quantum protocols for separable measurements with many parties}
\author{Scott M. Cohen}
\email{cohensm52@gmail.com}
\affiliation{Department of Physics, Portland State University,
Portland, Oregon 97201}

\begin{abstract}
In a recent paper \cite{mySEPvsLOCC}, we showed how to construct a quantum protocol for implementing a bipartite, separable quantum measurement using only local operations on subsystems and classical communication between parties (LOCC) within any fixed number of rounds of communication, whenever such a protocol exists. Here, we generalize that construction to one that applies for any number of parties. One important observation is that the construction automatically determines the ordering of the parties' measurements, overcoming a significant apparent difficulty in designing protocols for more than two parties. We also present various other results about LOCC, including showing that if, in any given measurement operator of the separable measurement under consideration, the local parts for two different parties are rank-$1$ operators that are not repeated in any other measurement operator of the measurement, then this separable measurement cannot be exactly implemented by LOCC in any finite number of rounds.
\end{abstract}

\date{\today}

\maketitle

\section{Introduction}\label{sec1}
It has long been a goal in quantum information theory to understand what can be accomplished by spatially separated parties sharing a multipartite quantum system when they cannot bring the parts together in one laboratory. In this case, the parties are restricted to local operations on subsystems and classical communication between parties (LOCC), and numerous results toward understanding LOCC have been achieved over the years \cite{Walgate,BennettPurifyTele,BennettConcentrate,CiracEnt,Nielsen,Anselmi,HillerySecret,DBerry,ourNLU,Chefles}. It is easily shown that the class of LOCC is a subset of the class of separable operations \cite{Rains}, and many studies of separable operations have been pursued as a means of understanding LOCC \cite{Gheorghiu2,Chefles,ourLocalCloning,ChitambarDuan,StahlkeU}. An important discovery was found in \cite{Bennett9}, where it was shown that a certain set of product states could not be perfectly distinguished using LOCC. This was the first demonstration that the class of separable operations is strictly greater than LOCC, and many other examples soon followed \cite{IBM_CMP,NisetCerf,AlonLovasz}.

In a recent paper \cite{mySEPvsLOCC}, we described a construction of LOCC protocols for implementing separable quantum measurements acting on a bipartite system. We also proved that this construction provides such a protocol whenever one exists in any fixed, finite, but arbitrary, number of rounds of communication, and will also determine when no such protocol exists. The proof was given for bipartite systems, and the main purpose of the present paper is to extend the construction, and the proof that it accomplishes what has just been claimed in the preceding sentence, to the case of any number of parties.

Let us begin by recalling what LOCC involves. We assume the parties have agreed in advance upon a protocol that they will all follow. One of the parties, say party $1$ whose system is described by states in Hilbert space $\HC_1$, starts by locally performing a generalized measurement \cite{Kraus} with outcomes corresponding to Kraus operators $K_{i_1}^{(;1)}$ (the index following the semi-colon in the superscript indicates on which one of the parties' subsystems this Kraus operator acts). If the initial multipartite state was $\ket{\Psi}$, then the state following this first measurement, given outcome $i_1$, will be $(K_{i_1}^{(;1)}\otimes I_2\otimes\ldots\otimes I_P)\ket{\Psi}$, where $I_\alpha$ is the identity operator on $\HC_\alpha$. This party broadcasts her outcome $i_1$ to the other parties, who according to the agreed upon protocol, all know which of them (call this party $2$) is to measure next. Party $2$ then performs a measurement with outcome $i_2$, described by $K_{i_2}^{(i_1;2)}$ acting on $\HC_2$ and conditioned on Alice's outcome $i_1$, after which he broadcasts his outcome $i_2$ to all the others. The next party to measure will be $\alpha$ (which could be party $1$ again), performing $K^{(i_1,i_2;\alpha)}_{i_3}$, and they may continue in this way for an arbitrary number of rounds. From the fact that the probabilities of outcomes obtained at each stage must always sum to unity, one has that for each and every $n$,
\begin{eqnarray}\label{eqn2001}
	I_\alpha = \sum_{i_n}K^{(\SC^\alpha_n)\dagger}_{i_n}K^{(\SC^\alpha_n)}_{i_n},
\end{eqnarray}
where $\SC^\alpha_n$ is a collection of indices $\SC^\alpha_n=\{i_1,i_2,\cdots,i_{n\!-\!1};\beta\}$ indicating all outcomes obtained in earlier measurements. The set of outcomes $i_1,\ldots,i_{n\!-\!1}$ determines $\beta$ from the agreed upon protocol, so that inclusion of this last index, $\beta$, in the collection is not strictly necessary. Nonetheless, we include it here for clarity. When $\beta\ne\alpha$ we define $K_{i_n}^{(\SC^\alpha_n)}=I_\alpha$ (this simply reflects the fact that party $\alpha$ does nothing when party $\beta$ is measuring), the sum on the right has only this single term, and \eqref{eqn2001} becomes trivial.

Every measurement represents a branching to one of multiple possibilities, so it is clear that the entire LOCC protocol can be represented as a tree. Each node corresponds to an outcome of one party's measurement, an edge connects that node to another node on its left, this latter node representing the immediately preceding outcome of a different party's measurement. We have found it useful \cite{mySEPvsLOCC} to label each node by a positive operator obtained by multiplying the product of all Kraus operators that party has performed leading up to and including the outcome that node represents, by its Hermitian conjugate. That is, the node reached by the set of outcomes, $(\SC^\alpha_n,i_n)$, will be labeled by
\begin{align}\label{eqn2002}
\KC_{i_n}^{(\SC^\alpha_n)}=K_{i_1}^{(\SC^\alpha_1)\dag}K_{i_2}^{(\SC^\alpha_2)\dag}\ldots K_{i_{n\!-\!1}}^{(\SC^\alpha_{n\!-\!1})\dag}K_{i_n}^{(\SC^\alpha_n)\dag}K_{i_n}^{(\SC^\alpha_n)}K_{i_{n\!-\!1}}^{(\SC^\alpha_{n\!-\!1})}\ldots K_{i_2}^{(\SC^\alpha_2)}K_{i_1}^{(\SC^\alpha_1)}.
\end{align}
Note that in this product of Kraus operators, there is one for each round leading up to this node, but many of these operators will be the identity, as parties other than $\alpha$ will have measured at that round along this branch of the tree. This approach effectively labels each node by $P$ positive operators, one for each party. However, since in moving from one node to the next along any branch, the only operator that changes is for the party which measured at that round, it is only this party's operators that will be displayed in the figures. The other operators can be deduced by looking at downstream nodes (those further from the leaves and toward the root of the tree), locating the nearest one that is labeled for the party in question.

As observed in \cite{mySEPvsLOCC}, \eqref{eqn2001} and \eqref{eqn2002} tell us that
\begin{align}\label{eqn2003}
\sum_{i_n}\KC_{i_n}^{(\SC^\alpha_n)}=\KC_{i_{n\!-\!1}}^{(\SC^\alpha_{i_{n\!-\!1}})}.
\end{align}
Recall that if it was not party $\alpha$ that measured at the $n$th round for the given branch, $K_{i_n}^{(\SC^\alpha_n)}=I_\alpha$, which when inserted into \eqref{eqn2002} and noting that the sum on the left of \eqref{eqn2003} then includes only a single term, we see that the latter equation is trivially satisfied.

Using these ideas in \cite{mySEPvsLOCC}, we showed how to construct an LOCC protocol in any finite number $R$ rounds whenever such a protocol exists for a given separable measurement on two parties. Here, we consider an arbitrary number of parties, $P$. By a separable measurement we mean a fixed collection $\{\widehat K_j^{(1)}\otimes\ldots\otimes\widehat K_j^{(P)}\}_{j=1}^N$ of distinct product Kraus operators for which there exists a set of positive coefficients, $\{\widehat w_j\}$, such that
\begin{eqnarray}\label{eqn2004}
	I_1\otimes I_2\ldots\otimes I_P = \sum_{j} \widehat w_j\widehat\KC_j^{(1)}\otimes\ldots\otimes\widehat \KC_j^{(P)},
\end{eqnarray}
where $\widehat \KC_j^{(\alpha)}=\widehat K_j^{(\alpha)\dagger} \widehat K_j^{(\alpha)}$. We emphasize that our definition of a measurement is in terms of the set of Kraus operators $\widehat K_j^{(\alpha)}$, and not just the positive operators $\widehat \KC_j^{(\alpha)}$, and that there may be more than one set of coefficients, $\widehat w_j$, such that \eqref{eqn2004} is satisfied. The construction begins by representing each positive operator, $\widehat \KC_j=\widehat\KC_j^{(1)}\otimes\ldots\otimes\widehat \KC_j^{(P)}$, as a linear (non-branching) tree of $P$ nodes, and then proceeds at each stage by merging smaller trees to create larger ones.

The remainder of the paper is organized as follows: In the next section, we recall how the construction works for the case of two parties. Then, in Section~\ref{sec3}, we discuss how to generalize the notion of merging trees from the bipartite case described in \cite{mySEPvsLOCC} to the case of more than two parties and then present a series of results, including a powerful sufficient condition that a separable measurement cannot be exactly implemented by LOCC. In Section~\ref{sec4}, we discuss how the construction automatically determines how to choose the order in which the parties measure, and in Section~\ref{sec5} we consider cases where one party flips a coin to decide which party will measure next. Following this, we offer our conclusions. In Appendix~\ref{secA1}, a detailed algorithm is presented for our construction, valid for any number of parties, and in Appendix~\ref{secA2} it is shown that this construction will provide an LOCC protocol in $R$ rounds whenever one exists, for any finite but arbitrary $R$.

Before proceeding, let us make a few comments about notation. We will refer to leaf and root nodes of a tree. There will always be $P$ root nodes at the left of every tree, only the right-most one will have more than one edge emerging from it. The leaves will be the terminal nodes along any branch at the other end of the tree. Upstream will mean toward the leaves, while downstream will mean toward the roots. For any given product operator, say $\widehat\KC_j=\widehat \KC_j^{(1)}\otimes\ldots\otimes\widehat \KC_j^{(P)}$, a `local' operator from $\widehat\KC_j$ will mean one of the $\widehat\KC_j^{(\alpha)}$, that operator being local to party $\alpha$. 

In the following, we will discuss the construction in general terms and also provide examples. In the examples, we will find it easier to denote parties as $A,B,C$, etc. and local operators as $\AC_j,\BC_j$, etc. (in particular, this notation is a bit less cumbersome in the figures). For general theorems and other results, on the other hand, we will use the notation appearing above, where local operators are denoted by $\widehat\KC_j^{(\alpha)}$, the superscript denoting to which party the operator is local.


\section{The construction in the bipartite case}\label{sec2}
We first review the case of two parties, for which one may assume the parties alternate their measurements. Let us denote the two parties by $A$ and $B$. Each measurement by the first party is followed by a measurement by the second, which is followed by the first party measuring again, and so on until the protocol is completed. We illustrate the construction of LOCC protocols by recalling example 4 of \cite{mySEPvsLOCC}, a separable measurement consisting of a set of five product operators $\{\widehat \AC_{j}\otimes\widehat \BC_{j}\}$ satisfying the constraints
\begin{align}\label{eqn10}
	\widehat \BC_1&=\widehat \BC_2=\widehat \BC_3\notag\\
	\widehat \BC_5&=\widehat \BC_1+\widehat \BC_4\notag\\
	I_B&=\widehat \BC_3+\widehat \BC_5\notag\\
	\widehat \AC_4&=\widehat \AC_1+\widehat \AC_2\notag\\
	I_A&=\widehat \AC_3=\widehat \AC_4+\widehat \AC_5,
\end{align}
and finally that there are no other linear constraints satisfied by these operators. 

\begin{figure}
\includegraphics{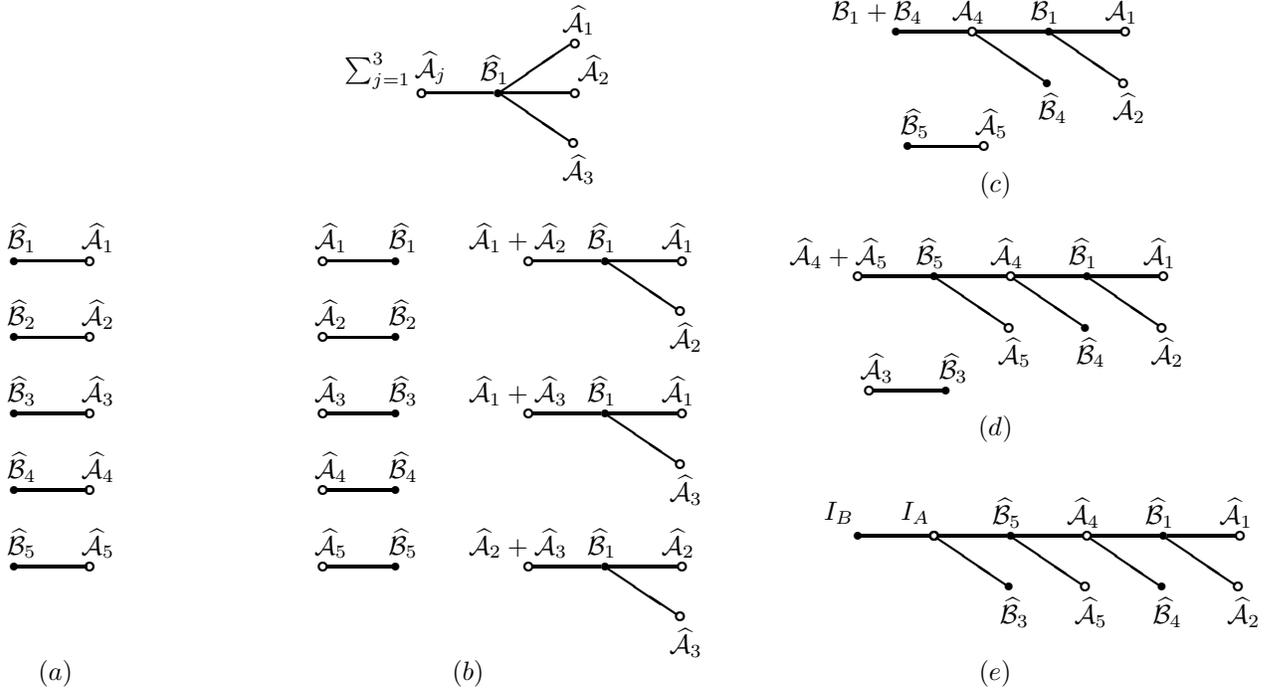}
\caption{\label{fig1}Step-by-step construction of the LOCC protocol for the measurement of Eq.~\eqref{eqn10}: (a) the five two-level trees representing the five $\widehat\AC_j\otimes\widehat\BC_j$, which are the starting point for the construction; (b) after the first step, we keep all the original two-level trees and add four more trees corresponding to the four different ways to merge the three proportional $B$-nodes, $\widehat\BC_1,\widehat\BC_2,\widehat\BC_3$; (c) after the second step we still have all two and three-level trees that are shown in (b) (these trees are not shown here, except for that representing $\widehat\AC_5\otimes\widehat\BC_5$, which will be used in the next step), along with one additional tree constructed from merging two trees from (b), the one that had $\widehat \AC_4$ as its left-most node and the one that had $\widehat \AC_1+\widehat \AC_2$ as its left-most node; (d) in the third step, we have one tree formed by merging the two trees depicted in (c), and a second tree that represents $\widehat\AC_3\otimes\widehat\BC_3$; (e) the two trees in (d) have been merged to form the final tree, representing the full LOCC protocol.}
\end{figure}

The construction begins by representing each of the five product operators by a two-node tree, as shown in part (a) of Fig.~\ref{fig1}. The next step is to identify all sets of $A$-nodes that are equal to each other, but from \eqref{eqn10} we see that there are none. So we look for sets of equal $B$-nodes, identifying $\widehat \BC_1,\widehat \BC_2,\widehat \BC_3$, and merge these in all possible ways ($\widehat \BC_1,\widehat \BC_2$; $\widehat \BC_1,\widehat \BC_3$; $\widehat \BC_2,\widehat \BC_3$; and $\widehat \BC_1, \widehat \BC_2,\widehat \BC_3$). This is followed by attaching a new $A$-node to the left, and labeling that node by the appropriate (according to \eqref{eqn2003}) sum of the $\widehat \AC_j$ operators ($\widehat \AC_1+\widehat \AC_2$ when merging $\widehat \BC_1,\widehat \BC_2$, etc.). These steps are shown in part (b) of the figure.  Since $\widehat \AC_1+\widehat \AC_2=\widehat \AC_4$, we merge the $\widehat\AC_4$ node of the $\widehat\AC_4\otimes\widehat\BC_4$ tree to the $\widehat \AC_1+\widehat \AC_2$ node in the tree from (b) that was obtained from merging $\widehat \BC_1,\widehat \BC_2$, see part (c) of the figure. The added left-most $B$-node is then labeled as $\widehat \BC_1+\widehat \BC_4$, which is the sum of those $B$-nodes immediately upstream from any given $A$-node that is itself immediately upstream of this left-most $B$-node (here, the only such $A$-node is that labeled $\widehat\AC_4$). Since $\widehat \BC_1+\widehat \BC_4=\widehat \BC_5$, the two-node tree representing $\{\widehat \AC_{5}\otimes\widehat \BC_{5}\}$ can be merged to the latter tree, and so on until the construction is completed when each $\{\widehat \AC_{j}\otimes\widehat \BC_{j}\}$ is included in a tree at least once, with the $P=2$ left-most nodes all being the identity operator for one of the parties, see part (e) of the figure.

With this warm-up exercise completed, we now turn to the question of merging trees when there are more than two parties.

\section{Merging trees with more than two parties}\label{sec3}
Each tree corresponds to a protocol of successive measurements by the parties. The merger of a set of trees combines those protocols into a single protocol by adding one additional, and earlier, round, this round being a measurement directly determined by the trees that were merged. The new tree must be consistent with each of the set that has been merged, in the sense that it preserves the protocols represented by those trees. For this to be the case, each of the nodes that are merged to each other must be equal (this is because for any two unequal nodes that are merged into one, the resulting node must then be labeled by an operator that is necessarily unequal to at least one of the operators that had previously labeled those two merged nodes, and therefore cannot satisfy \eqref{eqn2003} for both of the branches, one branch for each of the trees that were merged). Then, following this merger, one adds new nodes and labels them so as to be consistent with \eqref{eqn2003}. How one follows this procedure when there are only two parties is clear, merging a single set of nodes, one node from each of the trees that are merged, all those nodes being those of a given party. When the number of parties exceeds two, however, it is not immediately obvious when a given merger of trees will be allowed, or indeed, even in what way a merger can take place. Does one merge only a single set of nodes into one, as is done for the bipartite case? Or perhaps one should merge $P-1$ sets of nodes, one for each party, each set merged into a single node. Is there only one way this can be done, or can one merge some intermediate number of sets, as well?

\begin{figure}
\includegraphics{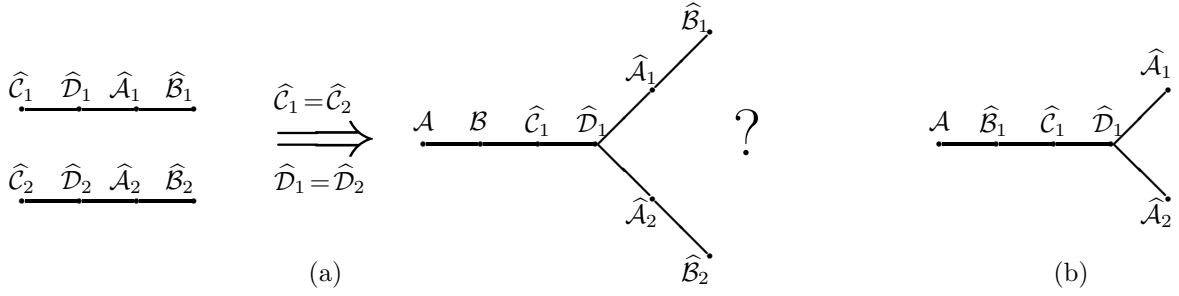}
\caption{\label{fig2}Can two four-party trees be merged when only two pairs of nodes are equal to each other? As discussed in the text, the merged tree at the right of (a) is not a valid LOCC protocol when $\widehat\BC_1\ne\widehat\BC_2$. On the other hand, when $\widehat\BC_1=\widehat\BC_2=\BC$, the tree on the right of (a) is then a valid LOCC protocol, and it is equivalent to the tree shown further to the right, in (b).}
\end{figure}

We need to determine under what conditions two (or more) trees can be merged when there are more than two parties. Consider as an example, the two four-party trees shown at the left of part (a) in Fig.~\ref{fig2} and representing the two product operators, $\widehat\AC_1\otimes\widehat\BC_1\otimes\widehat\CC_1\otimes\widehat\DC_1$ and $\widehat\AC_2\otimes\widehat\BC_2\otimes\widehat\CC_2\otimes\widehat\DC_2$. Can these trees be merged when $\widehat\CC_1=\widehat\CC_2$ and $\widehat\DC_1=\widehat\DC_2$, but $\widehat\AC_1\ne\widehat\AC_2$ and $\widehat\BC_1\ne\widehat\BC_2$? If so, then one obtains the merged tree shown at the right of (a) in the figure (or else a similar tree with the position of the $A$ and $B$ nodes swapped). This tree represents a two-outcome measurement by $A$, with each of these two outcomes being followed by a one-outcome measurement by $B$. Measurements having only one outcome must in fact be unitary operations, which implies by \eqref{eqn2003} that $\widehat\BC_1=\BC$ by following the upper branch, and $\widehat\BC_2=\BC$ by following the lower one (by \eqref{eqn2003}, we also have $\AC=\widehat\AC_1+\widehat\AC_2$). Therefore, this merger is possible only when $\widehat\BC_1=\widehat\BC_2$. When the latter equality holds, the merged tree is equivalent to the one shown in (b) of the figure.

While these arguments have been presented in the context of a specific example, in fact they are quite general. They certainly hold not only for four parties, but for any number $P$. Start with any separable measurement represented by $N$ linear trees of $P$ nodes, which may all be considered as root nodes. According to the above arguments, any valid merger of a subset of these trees must be done by merging them at $P-1$ of their root nodes, leaving a new tree that branches to those $P$th nodes, representing a measurement by that $P$th party. After each such merger, we add a new root node for that $P$th party to the newly formed tree, label it according to \eqref{eqn2003}, and then proceed to look for additional ways to merge trees. Since the arguments of the previous paragraph may be applied equally well to these new trees, given that those arguments are not altered by the presence of additional branches but only require consideration of the $P$ root nodes, we have the following lemma.
\begin{lem4}\label{lem4}
For the construction of all possible LOCC protocols (represented as trees) from any given separable measurement on $P$ parties, it is sufficient to consider only mergers that merge $P-1$ sets of root nodes (one for each party), and within each set, all nodes must be equal to each other.
\end{lem4}

Notice that the equivalence between the tree in (b) and the right-hand tree in (a) of Fig.~\ref{fig2} is a special case of the following simple, and rather obvious, observation. If, at any point other than at the very end of an LOCC protocol, any party performs a unitary operation, that party can just as well include that unitary operation as part of their next measurement. That is, if for party $\alpha$, $K_{1}^{(\SC^\alpha_n)}$ is equal to a unitary operator (and is therefore the only outcome of that particular measurement), and that party's next measurement along any given branch is defined by Kraus operators, $K_{i_m}^{(\SC^\alpha_m)}$ ($m>n$), then $\alpha$ may just as well skip that measurement $n$ and instead do a measurement with Kraus operators $K_{i_m}^{(\SC^\alpha_m)}K_{1}^{(\SC^\alpha_n)}$ at round $m$. There will be no difference whatsoever between the latter protocol and the former, and we therefore have the following lemma.
\begin{lem1}\label{lem1}
For the construction of all possible LOCC protocols for any given separable measurement, it is sufficient to consider only those trees for which every node (excluding leaf nodes, which terminate the protocol) branches to at least two other nodes, representing measurements that have two or more outcomes.
\end{lem1}

Clearly if the original set of product operators, which define the separable measurement to be implemented, is such that no two of them share $P-1$ pairs of proportional local operators, one pair for each of $P-1$ of the parties, then no merger of these trees is possible. This leaves us with no way to even begin to construct a full LOCC tree for this measurement. This idea can be generalized to give us the following theorem, in which we use the notion of a \emph{non-trivial} intersection of convex cones, meaning an intersection of cones other than the point at the origin (the zero operator). 
\begin{thm1}\label{thm1}
Given a separable measurement whose Kraus operators correspond to a set of positive product operators $\{\widehat\KC_j\}_{j=1}^N$ acting on $P\ge2$ parties, consider any partition of this set into two non-empty subsets, $S_1$ and $S_2$. Define the convex cones generated by each set of local operators within each of these two subsets as $C_1^{(\alpha)}$ and $C_2^{(\alpha)}$. If for any two parties, $\alpha\ne\beta$, the pair of cones $\{C_1^{(\alpha)},C_2^{(\alpha)}\}$ has no non-trivial intersection, and if the same is also true for $\{C_1^{(\beta)},C_2^{(\beta)}\}$, then no finite-round LOCC protocol exists for the exact implementation of this separable measurement.
\end{thm1}
\proof Let party $\alpha$ be $A$ with local operators $\widehat\AC_j$, and party $\beta$ be $B$ with local operators $\widehat\BC_j$. Recall that an LOCC tree for the separable measurement must merge all the  $\widehat\KC_j$ trees into a single tree. 

Consider the first step in our construction for building an LOCC protocol, which starts by merging the original $\widehat\KC_j$ trees. No one of these trees from $S_1$ can be merged to one from $S_2$, because (in order to have $P-1$ matching pairs, necessary according to Lemma~\ref{lem4}) such a merger would require either that the corresponding $\widehat\AC_j$ from $S_1$ is proportional to that from $S_2$, or that the corresponding $\widehat\BC_j$ from $S_1$ is proportional to that from $S_2$, both of which are false. On the contrary, at this step the only mergers involve $\widehat\KC_j$ that are all from $S_1$, or else $\widehat\KC_j$ that are all from $S_2$. After each merger a new node is added to the left of the merged tree. If that new node is an $A$-node, it will be labeled by a positive, linear combination of operators, all of which are from $S_1$ (or $S_2$), so this label will be an operator that lies in $C_1^{(A)}$ (or $C_2^{(A)}$), and similarly if the new node is a $B$ node. If, on the other hand, the new added node is not an $A$-node or a $B$-node, the root $A$ (and $B$) node for the merged tree is the same $\widehat\AC_j$ (and $\widehat\BC_j$) from the original $\widehat\KC_j$ trees that were merged. Clearly, these new trees just constructed satisfy the same conditions as do the original trees with regard to the partition into $S_1$ and $S_2$, that the local $A$ and $B$ operators labeling the root nodes on each of these trees still lie either in $C_1^{(A)}$ and $C_1^{(B)}$, or in $C_2^{(A)}$ and $C_2^{(B)}$, respectively. Therefore, the next mergers will produce new trees that also continue to satisfy this condition, and by extension, so will all mergers, no matter how many rounds are allowed. Hence, no tree is ever produced that includes operators $\widehat\KC_j$ from both $S_1$ and $S_2$, and so obviously no tree is produced that has merged all the $\widehat\KC_j$ into a single tree. This completes the proof.\endproof

Notice that the $\widehat\KC_j$ in $S_1$ cannot by themselves represent a complete measurement, and the same is also true of those in $S_2$. The former would require the existence of positive coefficients $\widehat w_{1j}$ such that $\sum_{j\in S_1}\widehat w_{1j}\widehat\KC_j=I$, the identity operator on the full multipartite space. However, given that the full set of $\widehat\KC_j$ (including both $S_1$ and $S_2$) is itself a complete measurement, this would mean that there also exist coefficients $\widehat w_{2j}$ such that $\sum_{j\in S_2}\widehat w_{2j}\widehat\KC_j=I$, and then it is not difficult to see that, for example, $I_A$ lies in both $C_1^{(A)}$ and in $C_2^{(A)}$, contradicting the conditions of the theorem.

Recall that an extreme ray of a convex cone is a ray that cannot be written as a positive linear combination of other rays in the cone (more precisely, no point on an extreme ray can be written as a positive linear combination of points in the cone, except if those points in the cone all lie on the extreme ray in question). In addition for any given separable measurement, let $\widehat\KC_n^{(\alpha)}$ for fixed $n$ be defined as a singular point, which generates a \emph{singular ray}, in the cone of party $\alpha$ if it is not proportional to any other $\widehat\KC_j^{(\alpha)},~j\ne n$. Then, we have the following immediate corollary to Theorem \ref{thm1}.
\begin{cor1}\label{cor1}
Consider any two parties, say $A$ and $B$, and any single product operator of the full separable measurement, say $\widehat\KC_n$ with local parts $\widehat \AC_n,\widehat \BC_n$ for those two parties. If $\widehat \AC_n$ is a singular extreme ray of the full cone of all $\widehat \AC_j$ operators from the full separable measurement, and if $\widehat \BC_n$ is a singular extreme ray of the full cone of all $\widehat \BC_j$ operators, then this separable measurement cannot be exactly implemented by any finite-round LOCC protocol.
\end{cor1}
\proof Define $S_1=\{\KC_n\}$ and let $S_2$ be the complement of $S_1$ in the full set of operators $\{\KC_j\}_{j=1}^N$ from the separable measurement. This corollary then follows immediately as a direct application of Theorem~\ref{thm1}.\endproof

In particular, if $\widehat \AC_n$ and $\widehat \BC_n$ are both rank-$1$ positive operators (which are extreme rays of the full space of positive operators and so are necessarily extreme in any cone lying within that space), neither of which are proportional to any of the other $\widehat \AC_j$ or $\widehat \BC_j$, respectively, then the separable measurement cannot be exactly implemented by finite-round LOCC. This gives a simple way to recognize (or construct) separable measurements that are not LOCC. One such example is the measurement that led to the discovery that the class of separable operations is strictly larger than the class of LOCC \cite{Bennett9}. For example, the latter measurement includes the operator $\widehat\KC_1=\dyad{1}{1}\otimes\dyad{1}{1}$, and the rank-$1$ local operator $\dyad{1}{1}$ does not appear in any other $\widehat\KC_j$ for either of the two parties. Therefore, this measurement satisfies the conditions of Corollary~\ref{cor1}.

\section{Determining the order of the parties' measurements}\label{sec4}
When there are only two parties, there is really no question as to the ordering of the parties' measurements. The only thing they can do is alternate, the second party following the first following the second, and so on. There is no need to consider that one party will follow their own measurement with another of their own, not even when this is the case for only some, and not all, outcomes of the first measurement. The reason is fairly obvious, that if an outcome implementing Kraus operator $A_1$ is followed by the same party measuring immediately again with Kraus operators $A_j^{(1)}$, that party might as well have omitted $A_1$ from the first measurement and included the set $\{A_j^{(1)}A_1\}$ in its place. There is effectively no difference between the two options.

When there are three or more parties, on the other hand, the number of possible sequences of ordering their measurements grows without bound as the number of rounds increases. So in addition to all the complications of designing an LOCC protocol for a given separable measurement already present in the case of two parties, there is the added question of deciding the ordering of the parties when there are more than two. Therefore, we believe the following result is not insignificant.
\begin{lem2}\label{lem2}
The construction of \cite{mySEPvsLOCC}, as modified here for more than two parties (see Appendix~\ref{secA1}), automatically determines which ordering(s) of the parties will yield a valid LOCC protocol.
\end{lem2}
\proof The argument is really quite simple. According to Lemma~\ref{lem4}, the decision as to which trees can be merged in following the construction of \cite{mySEPvsLOCC} is determined by whether or not the trees have $P-1$ root nodes that match. By `matching', we mean that the nodes are equal, which in the way we proceed with the algorithm (see Appendix~\ref{secA1}), means that each of the $P-1$ convex cones on one tree, one for each party's root node, intersects with that party's corresponding convex cones for the nodes on the other trees. Now, the root nodes (of which there are $P$ in every tree at any point in the construction) are effectively place-holders, as they do not themselves represent measurements, whereas the protocol being constructed is defined in terms of the sequence of measurements that are performed. Therefore, the order of the $P$ root nodes is irrelevant, and they can be re-ordered arbitrarily.

Each merger of two or more trees merges $P-1$ of the $P$ root nodes in those trees, which leaves one party's root node from each of those merged trees (the same party for all those trees) branching from the right-most root in the resulting, merged tree. This corresponds to a measurement by that $P$th party, that measurement having a number of outcomes equal to the number of trees that were including in that particular merger --- one outcome, one branch, from each of the trees that were merged. If exactly $P-1$ root nodes in one tree match their counterparts in another tree, there is no ambiguity as to how those two trees can be merged: they can only be merged in a way that corresponds to a measurement by that $P$th party. If, on the other hand, all $P$ root nodes in one tree match those in another, then there are $P$ ways those two trees can be merged, each corresponding to a measurement by a different one of those $P$ parties. By considering all of these, which is done in our algorithm (see step \ref{enum0} of the algorithm in Appendix~\ref{secA1}), we will find all possible orderings of the parties' measurements that produce valid LOCC protocols.\endproof

The case where all $P$ root nodes are matching is special and is considered further in the next section. One other point worth noting is that by considering all possible orderings of the root nodes, it is possible one may merge trees in a way that produces a measurement by a given party immediately following a measurement by that same party. This is not a problem, as it still yields a valid protocol, and as is argued in the beginning of this section, this protocol can be replaced by an equivalent one that incorporates both those measurements by that particular party into a single measurement. A bit of thought reveals that the latter measurement will also be constructed by our algorithm, so one may omit consideration of all trees that include successive local measurements by the same party.

\section{Flipping Coins}\label{sec5}
\begin{figure}
\includegraphics[scale=.75]{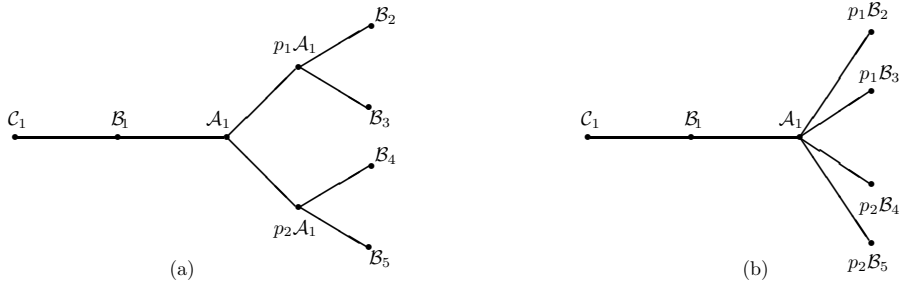}
\caption{\label{fig3}(a) This tree represents a measurement that can be thought of as $A$ flipping a biased coin to determine which branch of the protocol to follow. According to Eq.~\eqref{eqn2003}, we must have $p_1+p_2=1$ and $\BC_2+\BC_3=\BC_1=\BC_4+\BC_5$. Given these relations, we see that the tree in (b) also satisfies \eqref{eqn2003}, which therefore yields a valid LOCC protocol, as well. In fact, the trees in (a) and (b) yield protocols that implement the exact same separable measurement, corresponding to positive operators $p_1\AC_1\otimes\BC_2\otimes\CC_1$, etc., and this will remain true even if some or all of the $\BC_j$ are followed by additional rounds of measurements. One can view the protocol of (b) as $B$ flipping that coin instead of $A$, and then performing the appropriate measurement as decided by that coin flip.}
\end{figure}
When all $P$ root nodes are matching for two or more trees, we can merge these trees at $P-1$ of those root nodes, leaving those $P$th nodes from the merged trees as the outcomes of a measurement by that last party. Since all $P$ root nodes of the original trees were matching, the nodes corresponding to those outcomes are represented by positive sums of positive operators whose convex cones intersect with one another. This leaves open the possibility that this particular measurement is one that has two or more outcomes that are proportional to each other. When this is the case, it is effectively the same as having that party combine those outcomes into a single outcome, and when that outcome is obtained in the measurement, the given party flips a generally biased, many-sided coin to determine which of the original branches of the protocol to follow; in the example of Fig.~\ref{fig3}(a), $A$ can flip a biased coin and conditioned on whether it is heads or tails, then tell $B$ which of the two measurements, $\{\BC_2,\BC_3\}$ or $\{\BC_4,\BC_5\}$, he should perform. There is nothing wrong with such measurements, which pose no problem for our purpose of constructing LOCC protocols. In fact, they are only relevant when $P>2$. The reason is that if the measurements following each of those proportional outcomes are all performed by the same party (which they necessarily are if $P=2$), say $B$ as in Fig.~\ref{fig3}(a), then it is obvious that those proportional outcomes can be combined into a single outcome followed by a single measurement by $B$, the latter measurement including all the outcomes of the collection of $B$'s measurements that had followed in the original implementation, see Fig.~\ref{fig3}(b). Of course, the protocol of Fig.~\ref{fig3}(b) is equivalent to one where $B$ flips that coin instead of $A$ and then performs the appropriate measurement as decided by that coin flip. This just illustrates the obvious fact that it makes no difference which one of them has the coin. 

\begin{figure}
\includegraphics[scale=.75]{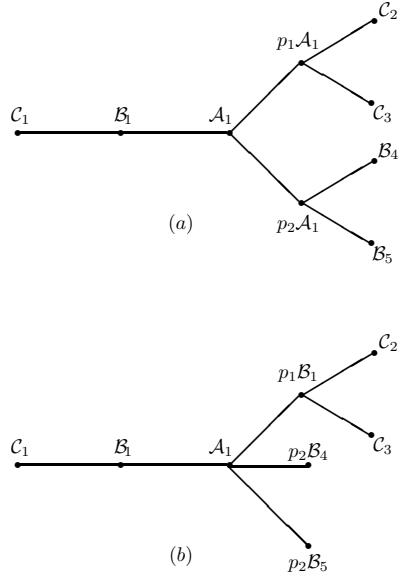}
\caption{\label{fig4}(a) Party $A$ flips a biased coin and then tells $B$ and $C$ which of them is to measure next. (b) Party $B$ flips the coin instead, telling $C$ (and $A$, if needed for subsequent rounds) whether $C$ should measure or if $B$, himself, will measure next. Since it makes no difference who flips the coin, the protocols of (a) and (b) yield identical results under all circumstances. Note that for (b), however, $B$ need not flip a coin, but can instead perform a single complete measurement having outcomes $p_1\BC_1,p_2\BC_4,p_2\BC_5$.}
\end{figure}

Having recognized that the equivalence between the protocols of (a) and (b) in  Fig.~\ref{fig3} is merely a reflection of the fact that either party can flip the coin, it should not be too surprising that the same conclusion holds even when the coin flip chooses between subsequent measurements by various different parties, a result stated in the following lemma.
\begin{lem3}\label{lem3}
For any LOCC protocol involving local measurements having two or more proportional outcomes, there is a corresponding LOCC protocol with no such measurements, but which implements the exact same separable measurement as the original protocol, including reproducing the same weights on each outcome of the separable measurement. In other words, there is never a need to use an LOCC protocol that includes measurements having multiple proportional outcomes (i.e., there is no need for the parties to flip coins).
\end{lem3}
\proof  While a detailed proof can be given, the easiest way to understand this result is from the idea that it doesn't matter who flips the coin(s). Consider the example in Fig.~\ref{fig4}. In (a) of the figure, party $A$ flips a biased coin and then tells $B$ and $C$ which of them is to measure next. In (b), party $B$ flips the coin instead, telling $C$ (and $A$, if needed for subsequent rounds) whether $C$ should measure or if $B$, himself, will measure next. The protocols obviously yield identical results, and will do so regardless of what measurements $B$ or $C$ perform following the coin flip, and even if the latter measurements are followed by a number of subsequent rounds. Furthermore, the tree in (b) need not be interpreted as a protocol with $B$ flipping a coin, $B$ can instead perform a single complete measurement having outcomes $p_1\BC_2,p_2\BC_4,p_2\BC_5$. The same conclusion (that there is no need for the parties to search their pockets for a spare coin) can be reached no matter what form the latter measurement by $B$ takes, and also if the outcomes of that measurement are followed by additional rounds of measurements.

\begin{figure}
\includegraphics[scale=.75]{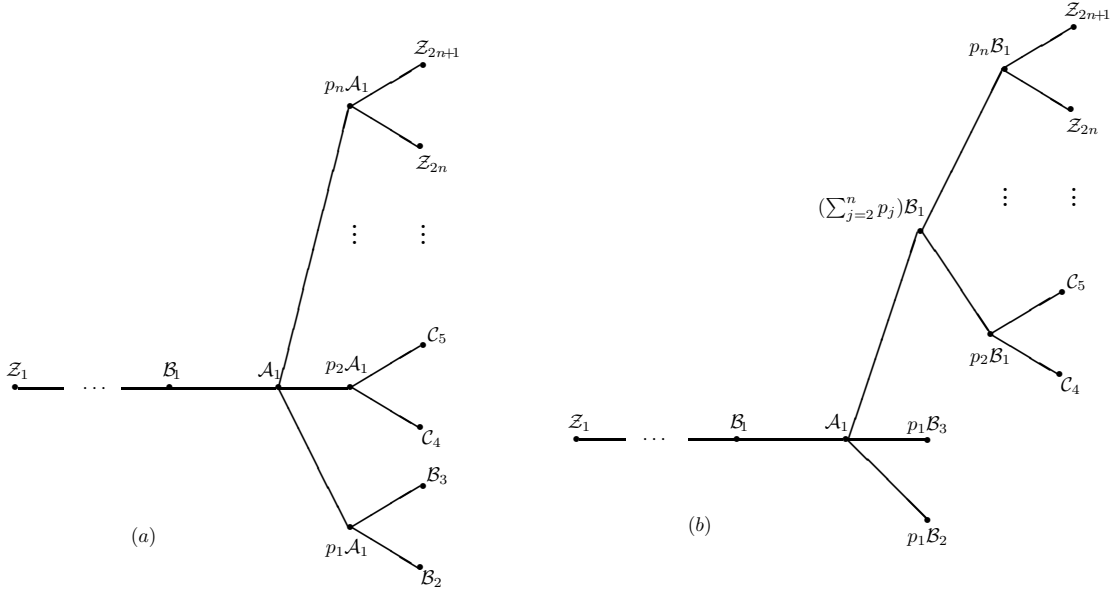}
\caption{\label{fig5}(a) Party $A$ decides who is to measure next by flipping a coin having more than two sides. (b) The first step in replacing $A$'s many-sided coin flip by a succession of two-sided coins being flipped by the other parties. First, as shown in (b), $B$ flips a two-sided coin to decide whether $B$ or someone else will measure next. If it is not $B$ to measure next, $C$ then flips a two-sided coin to decide who amongst the rest of them will measure (not shown). This process is continued to the end, leaving a protocol that includes no repeated (proportional) outcomes in any given local measurement.}
\end{figure}

In Fig.~\ref{fig5}(a), we have an example involving $A$ flipping a coin having more than two sides. Part (b) of this figure shows the first step toward replacing the many-sided coin flip by $A$, this first step involving $B$ flipping a two-sided coin. If this latter coin comes up tails (outcome 2), $B$ measures and informs everyone else of this fact ($B$ can also tell them the outcome of his measurement in case they need this information for subsequent rounds). If $B$'s coin comes up heads (outcome 1), he does not measure and just tells the rest that they need to decide who is to measure next. In the latter case, $C$ can next flip a two-sided coin, for one outcome of which she measures herself, for the other she does not, in either case communicating the necessary information to everyone else. Afterward, the next party flips a coin, then the next, and so on until they have included all the required possibilities. From this perspective, the statement of the lemma will clearly hold no matter what measurements each of the parties performs after $A$'s original coin-flip, and even if all their measurements are then followed by a number of subsequent rounds. In addition, one can easily replace those two-outcome coin flips with a single complete measurement by each of the parties, so there is no need for them to flip coins.\hspace{\stretch{1}}$\blacksquare$

\section{Conclusions}\label{conc}
To summarize, we have shown how to generalize the construction of LOCC protocols for a given separable measurement from the bipartite case described in \cite{mySEPvsLOCC} to the case of more than two parties. We presented a series of results about the existence of such protocols, results that are valid for any number of parties, including a powerful sufficient condition that a separable measurement cannot be exactly implemented by LOCC, see Theorem~\ref{thm1}. We showed that our construction automatically determines how to choose the order in which the parties measure, overcoming a significant apparent difficulty in designing LOCC protocols when there are more than two parties. We also considered cases where one party flips a coin to decide which party will measure next, and showed that one can dispense with the use of coins altogether, leaving only protocols that never include local measurements having multiple repeated (proportional) outcomes. In Appendix~\ref{secA1}, a detailed algorithm is presented for our construction, valid for any number of parties, and in Appendix~\ref{secA2} we demonstrate that this construction will provide an LOCC protocol in $R$ rounds whenever such a protocol exists, for any finite but arbitrary $R$.

As the reader will have noted, our construction of LOCC trees from a given separable measurement is done without direct consideration of the Kraus operators $\widehat K_j$ for that measurement, but rather it is only necessary to consider the positive operators $\widehat K_j^\dag\widehat K_j$ corresponding to those Kraus operators. Thus, the question as to the existence of an LOCC protocol for a given separable measurement depends only on those positive operators and not on the Kraus operators, themselves. We therefore have the following result.
\begin{thm2}\label{thm2}
Given two separable measurements defined in terms of Kraus operators as $\{\widehat K_i\}_{i=1}^N$ and $\{\widehat K_j^\prime\}_{j=1}^{N^\prime}$, suppose the two yield the same set of distinct positive operators up to multiplicative factors. That is, every operator $\widehat\KC_i=\widehat K_i^\dag\widehat K_i$ is proportional to one of the $\widehat\KC_j^\prime=\widehat K_j^{\prime\dag}\widehat K_j^\prime$, and vice-versa. Then there exists an LOCC protocol exactly implementing the first separable measurement if and only if there exists an LOCC protocol exactly implementing the second separable measurement.
\end{thm2}
\noindent When an LOCC protocol exists for both measurements, the two protocols are exactly the same except that extra unitaries must be performed at the end. These unitaries $U_{ji}$ are determined by the relationship, $\widehat K_j^\prime=\sqrt{p_{ji}}U_{ji}\widehat K_i$, necessary for the positive operators to be proportional to each other. If, for a given one of the separable measurements, more than one Kraus operator yield the same $\widehat\KC_i$ up to a multiplicative factor, then one party can flip a coin (biased according to the different $p_{ji}$, for example) to determine which (product) unitary the parties need to implement when that $\widehat\KC_i$ is obtained. This may require one additional round of communication to share the outcome of that coin flip with the other parties, so they will know what local unitary they are to perform.

\noindent\textit{Acknowledgments} --- This work has been supported in part by the National Science Foundation through Grant No. 1205931.
\appendix
\section{The algorithm} \label{secA1}

We here give a detailed step-by-step algorithm for the construction of LOCC protocols from any given separable measurement corresponding to the set of positive product operators, $\widehat\KC_j=\widehat\KC_j^{(1)}\otimes\ldots\otimes\widehat\KC_j^{(P)},~j=1,\ldots,N_0$. This algorithm was provided in \cite{mySEPvsLOCC} for the case of two parties, and here it is modified to apply to any number $P$ of parties. At various steps along the way in this algorithm, we create multiple copies of certain trees that had previously been constructed. The point of this is to be sure we merge these trees to other ones in all ways it is possible for that tree to be merged. That is, one copy of a given tree is created for each possible way that tree can be merged to other trees, a different copy of that particular tree being used for each different merger. 

The algorithm involves the following step-by-step process.
\begin{enumerate}
  \item \label{enum0} Start with one $P$-node tree for each of the $\widehat\KC_j=\widehat\KC_j^{(1)}\otimes\ldots\otimes\widehat\KC_j^{(P)},~j=1,\ldots,N_0$, and include positive factors $\widehat q_{j1}^{(\alpha)}$ with each $\widehat\KC_j^{(\alpha)}$ -- that is, label each node as $\widehat q_{j1}^{(\alpha)}\widehat\KC_j^{(\alpha)}$; partition the collection of all these trees into equivalence classes, every tree in any given class having a fixed $P-1$ of their nodes all proportional to each other (for example, one class might have nodes for party $1$ from all the various trees proportional, all nodes for party $2$ proportional, etc., up to party $P-1$); arrange the $P$ nodes of each tree such that the chosen $P-1$ are to the left of the last, $P$th, one, and are all arranged in the same order, which may be chosen arbitrarily. Set $r=0$, which will serve as a counter for the depth of the trees.
  \item \label{enum10} WHILE $r<R$ ($R$ is the maximum number of rounds to be considered)
  \item \label{enum1} Increment $r$. For each equivalence class and for each subset in that class that contains at least one member: (a) CREATE one copy of each tree in the given subset; (b) MERGE left-most nodes of all these copies into $P-1$ nodes, one for each party save the $P$th, and relabel the coefficients $\widehat q_{jk}^{(\alpha)}$ with a unique value of the $k$ index for every different appearance of $\widehat\KC_j$ at the leaves within the merged tree, adjusting those $k$ indices throughout the rest of the tree to be consistent [according to \eqref{eqn2003}] with its leaf labels; (c) EXTEND the tree by attaching a new node to the left, and label that new node to obey \eqref{eqn2003}, this node being a node of the $P^\textrm{th}$ party, the one that is \textit{not} used to determine members of the given equivalence class; (d) RECORD all constraints that the merged nodes must be equal (at the first step, these will be of the form $\widehat q_{ik}^{(\alpha)}\widehat\KC_i^{(\alpha)}=\widehat q_{jk^\prime}^{(\alpha)}\widehat\KC_j^{(\alpha)}$ when node $\widehat\KC_i^{(\alpha)}$ is merged with node $\widehat\KC_j^{(\alpha)}$ and are constraints on the $\widehat q^{(\alpha)}$'s). Number the \emph{new} trees sequentially from $N_{r-1}+1$ to $N_r$, where $N_r$ is the total number of trees at this stage (including trees of all depths constructed so far). 
 [Note that we can ignore each equivalence class that is identical to one that was previously present at an earlier pass through this algorithm, as all trees that can be constructed from that equivalence class have already been constructed.]
  \item \label{enum2} FOR $m=N_{r-1}+1$ to $N_r$

  \item \label{enum4} IF the $m^{\textrm{th}}$ tree includes each of the $\widehat\KC_j$ at least once, then: For the collection of all constraints recorded in (multiple passes through) step \ref{enum1} for the $m^{\textrm{th}}$ tree, check to see if there exists a solution for the $\widehat q_{jk}^{(\alpha)}$, including constraints that the $P$ left-most nodes in the tree under consideration are labeled by $I_1,\ldots,I_P$ (we start this FOR loop at $N_{r-1}+1$ since all the earlier trees have already been examined; for $r=1$, we know that the first $N_0$ of the trees have only a single $\widehat\KC_j$, so cannot include each of them at least once). If such a solution exists, we are done, having identified an LOCC protocol, so exit and END this algorithm (or one can continue if the aim is to build all possible trees). 
    \item \label{enum44} END FOR ($m$)
\item \label{enum3}We now have an expanded set of trees with the $P$ left-most nodes labeled by sums of $\widehat q_{jk}^{(\alpha)}\widehat\KC_j^{(\alpha)}$. Consider the convex cones generated by the sets $\{\widehat\KC_j^{(\alpha)}\}$ appearing in each such sum for each party $\alpha$, and identify an equivalence class for each subset of these trees such that the associated convex cones share a common mutual intersection for some choice of $P-1$ of the parties (but the same $P-1$ for any given class; a given tree will generally be included in multiple equivalence classes). Each such set of $P-1$ intersections implies that the associated trees can be merged, which we will do next, so go back to step \ref{enum10} and repeat. However, we only need to look for \emph{new} equivalence classes, involving newly constructed trees (along with all previous ones), since we've already constructed all trees that derive from the other equivalence classes. If there are no new classes then no new trees can be constructed, which means since no previously constructed tree has been found to be LOCC, then no LOCC protocol exists for this measurement, no matter how many rounds are allowed. Therefore, exit and END this algorithm.
  \item END WHILE ($r$)
\end{enumerate}
Note that as we loop through the WHILE loop, we keep \emph{all} trees for the next round, including not just those constructed in the present round, but also those from all previous rounds (we also keep all constraints). This makes it possible for multiple trees of differing size and structure and such that several of them each include the same $\widehat\KC_j$ (or even several of the $\widehat\KC_j$ that are repeated in this way), to be merged together into a single tree. 

\section{Proof that the algorithm works}\label{secA2}
The proof in \cite{mySEPvsLOCC} that the algorithm builds an LOCC protocol in $R$ rounds for a given separable measurement whenever one exists was given only for the case of two parties. We therefore need to extend that proof to cover any number of parties. The proof for two parties involved three parts: 
\begin{enumerate}
\item An equivalence was demonstrated between LOCC trees and LOCC protocols.
\item It was shown that if, for a given separable measurement, a tree exists that includes a merger of two or more congruent sub-trees, then there also exists a tree for the same separable measurement that does not include any mergers of congruent sub-trees.
\item \label{item3} Starting from \textit{any} LOCC tree that does not include a merger of congruent sub-trees, and taking that tree apart in a step-by-step fashion, it was shown that the algorithm of Appendix~\ref{secA1} would build that tree, from which one then concludes that the construction provides an LOCC protocol in $R$ rounds whenever one exists.
\end{enumerate}
The first two items in this list were proven by considering each party separately, without reference to the other party, so those arguments do not depend on the number of parties and continue to hold for $P>2$.\footnote{We note here a correction for Eqs. (B8) and (B9) of \cite{mySEPvsLOCC}. Everywhere that $A_{i_m}^{(S_m)}$ appears in these two equations (which it does in various places within parentheses), it should be replaced by $\sqrt{\AC_{i_m}^{(S_m)}}$.} The last item, on the other hand, was demonstrated by considering trees in their entirety, and therefore could depend on $P$. We now argue that this last part of the proof can be easily altered to apply as well to any number of parties.

Item~\ref{item3} in the proof of the main theorem of \cite{mySEPvsLOCC} shows that the construction produces any LOCC tree by starting from an arbitrary such tree, cutting it into pieces of varying size --- starting with trees of depth two, which are cut from the leaf end; then trees of depth three, which are again cut from the leaf end; etc. --- and showing that each of these subtrees will be constructed by our algorithm. Since in the construction, we always have trees with $P$ root nodes, the depth of these trees will be at least $P$, so the construction will not build trees of depth less than $P$. So it may appear there are problems in the proof for $P>2$, at least in the beginning where we start by cutting off trees of depth two. However, this difficulty is easily circumvented by the following modification, more accurately following our construction, which builds trees that always have $P$ roots: After cutting trees of any depth from the leaf end, 
extend that tree by attaching $P-1$ root nodes at its left so that there is a root node for each party (the cut will always produce a tree with one root node, the node that is farthest to the left and from which all other nodes emerge). If the tree produced by the cut contains nodes for each of the parties, then label the added root nodes according to \eqref{eqn2003}, which insures this will match the labeling appearing on that party's first downstream node in the original tree, since the original tree satisfies \eqref{eqn2003} to begin with. For any party that has none of her nodes appearing in the tree produced by the cut, that party's root node should be labeled by the positive operator on that party's first node downstream from the edge that was cut in the original tree to produce the (sub)tree in question. This positive operator will necessarily be one of the local operators $\widehat\KC_j^{(\alpha)}$ defining the separable measurement implemented by the LOCC protocol --- this is true by definition, since this is precisely how one determines the separable measurement implemented by the LOCC protocol, by looking at the first $\alpha$-node (for each $\alpha$) downstream from any given leaf. Since re-ordering of the $P$ left-most nodes in any tree is immaterial, this allows all the arguments given in the proof for two parties to go through unchanged for more than two parties. Hence, we conclude that our construction will provide an LOCC protocol in $R$ rounds whenever one exists for any given separable measurement, no matter how many parties are involved. 

%

%
\end{document}